\documentclass[useAMS,usenatbib,usegraphicx]{mn2e}

\title[Polarization of Bright Stars]{The linear polarization of nearby bright stars measured at
the parts per million level}
\author[J. Bailey et al.]{Jeremy Bailey,$^1$\thanks{E-mail: j.bailey@unsw.edu.au (JB)} 
P.W.~Lucas,$^2$
J.H.~Hough$^2$ \\
$^{1}$School of Physics, University of New South Wales, NSW 2052, Australia\\
$^{2}$Centre for Astrophysics Research, Science \& Technology
Research Institute, University of Hertfordshire, Hatfield, AL10 9AB, UK}
\begin{document}

\date{Accepted 2010 Mar 6; Received 2010 Mar 5; in original form 2010 Jan 17 }

\pagerange{\pageref{firstpage}--\pageref{lastpage}} \pubyear{2010}

\maketitle

\label{firstpage}

\begin{abstract}  We report observations of the linear polarization of a sample of 49
nearby bright stars measured to sensitivities of between $\sim$1 and $\sim$4 $\times
10^{-6}$. The majority of stars in the sample show measurable polarization, but most
polarizations are small with 75\% of the stars having P $<$ 2 $\times10^{-5}$. 
Correlations of the polarization with distance and position,  indicate that most of the
polarization is of interstellar origin. Polarizations are small near the galactic pole and
larger at low galactic latitudes, and the polarization increases with distance. However,
the interstellar polarization is very much less than would be expected based on
polarization-distance relations for distant stars showing that the solar neighbourhood has
little interstellar dust. BS 3982 (Regulus) has a polarization of $\sim$ 37 $\times
10^{-6}$, which is most likely due to electron scattering in its rotationally flattened atmosphere. BS 7001 (Vega)
has polarization at a level of $\sim$ 17 $\times 10^{-6}$ which could be due to scattering
in its dust disk, but is also consistent with interstellar polarization in this direction.   
The highest polarization observed is that of BS 7405 ($\alpha$ Vul)
with a polarization of 0.13\% \end{abstract}

\begin{keywords}
polarization -- techniques: polarimetric -- ISM: magnetic fields.
\end{keywords}

\section{Introduction}

Linear polarization of starlight provides a powerful technique for investigating the nature of 
the interstellar medium. Interstellar dust particles aligned to the galactic magnetic field produce
interstellar polarization, which is one of the main sources of stellar linear polarization. Studies
of this polarization provide information on the dust distribution and magnetic field
structure \citep[e.g.][]{heiles96} and on the nature and size of the dust particles
\citep{whittet92,kim94}. 

Studies of the polarization of stars close to the Sun have been made by \citet{piirola77},
\citet{tinbergen82} and \citet{leroy93a,leroy93b,leroy99}. They found very little polarization in
neraby stars. \citet{leroy93b} found only 25 stars with definite polarization in a survey of 1000
stars within 50pc. Subsequent analysis showed that almost all of these polarized stars were
actually at greater distances when more accurate Hipparcos parallaxes became available
\citep{leroy99}, and that significant interstellar polarization became detectable at distances of
about 70pc in some directions and at 150pc in others. \citet{andersson06} have also reported
observations of polarization of southern hemisphere stars, that they attribute to the wall of
the Local Bubble at $\sim$100 pc distance. 

All of these studies used polarization measurements with accuracies of, at best, $\sim$10$^{-4}$ in
fractional polarization. Recently we have built and tested a new polarimeter, PlanetPol,
\citep{hough06} capable of measuring stellar linear polarization at the parts per million level.
Here we report observations of a sample of nearby bright stars measured to sensitivies of generally
better than 3 $\times$ 10$^{-6}$ and in some cases to better than 1 $\times 10^{-6}$ in fractional
polarization. This represents an improvement of a factor of 20 to 100 on previous measurements.

In addition to the use of polarization to probe the interstellar medium, it is of interest to
know at what level normal stars show intrinsic polarization. There is currently considerable
interest in using polarization to study extrasolar planets. Polarization can be used  to
directly detect  unresolved hot-Jupiter type planets \citep{seager00,lucas06,lucas09}, as a
differential technique to detect planets in imaging observations \citep{schmid05,keller06}, and
as a means of characterizing extrasolar planet atmospheres \citep{bailey07}. Significant
polarization from the host star could complicate such observations. In the case of the quiet Sun
direct observations by \citet{kemp87} show linear polarization of $<3 \times 10^{-7}$.
 However, more active stars
could show higher polarizations, and polarization might also result from exozodiacal disks
around the stars.

\begin{table*}
\caption{Properties of Sample Stars - The horizontal line marks the 17 hour division used in figure 4}
\begin{flushleft}
\begin{tabular}{llrlrrrrrr}
\hline
BS   & Other Names  & V & Spectral & Dist & RA & Dec  &  \multicolumn{2}{c}{Galactic} & $v \sin{i}$ \\
       &                       &  mag   &  Type    & (pc) &  hh:mm  &  dd:mm  &   Long & Lat   &  km$s^{-1}$ \\   \hline                
3982 & Regulus, $\alpha$ Leo & 1.35  &  B7V     & 23.8 &  10 08 &  +11 58 &  226.4 & 48.9  &  353 \\
4031 &                       & 3.44  &  F0III   & 79.6 &  10 16 &  +23 25 &  210.2 & 55.0  &  83 \\
4069 &                       & 3.07  &  M0III   & 76.3 &  10 22 &  +41 30 &  177.9 & 56.4  &   \\
4295 & Merak, $\beta$ Uma    & 2.35  &  A1V     & 24.3 &  11 01 &  +56 23 &  149.2 & 54.8  &  32  \\
4301 & Dubhe, $\alpha$ Uma   & 1.79  &  K0Iab   & 37.9 &  11 03 &  +61 45 &  142.8 & 51.0  &  $<$17 \\
4335 &                       & 3.01  &  K1III   & 45.0 &  11 09 &  +44 30 &  165.8 & 63.2  &  10 \\
4357 &                       & 2.56  &  A4V     & 17.7 &  11 14 &  +20 31 &  224.2 & 66.8  &  180  \\
4359 &                       & 3.32  &  A2V     & 54.5 &  11 14 &  +15 26 &  235.4 & 64.6  &  5 \\
4518 &                       & 3.71  &  K0.5IIIb & 60.1 & 11 46 &  +47 47 &  150.3 & 28.4  &  10 \\
4527 &                       & 4.54  &  A7V     & 69.4 &  11 48 &  +20 13 &  235.0 & 73.9  &   \\
4534 & $\beta$ Leo           & 2.14  &  A3V     & 11.1 &  11 49 &  +14 34 &  250.6 & 70.8  &  110 \\
4540 & $\beta$ Vir           & 3.61  &  F9V     & 10.9 &  11 51 &  +01 46 &  270.5 & 60.8  &  3 \\
4905 &                       & 1.76  &  A0p     & 24.8 &  12 54 &  +55 58 &  122.2 & 61.2  &  33 \\
4910 &                       & 3.38  &  M3III   & 62.1 &  12 56 &  +03 24 &  305.5 & 66.2  &   \\
4915 & $\alpha^2$ CVn        & 2.90  &  A0p     & 33.8 &  12 56 &  +38 19 &  118.3 & 78.8  &   \\
4932 &                       & 2.83  &  G8III   & 31.3 &  13 02 &  +10 58 &  312.3 & 73.6  &  8 \\
5054 & Mizar                 & 2.27  &  A2V     & 24.0 &  13 24 &  +54 56 &  113.1 & 61.6  &  13 \\
5191 &                       & 1.85  &  B3V     & 30.9 &  13 47 &  +49 19 &  100.7 & 65.3  &  226 \\
5235 &                       & 2.68  &  G0IV    & 11.3 &  13 55 &  +18 24 &    5.3 & 73.0  &  18 \\
5340 & Arcturus, $\alpha$ Boo & $-$0.04 & K1.5III & 11.3 &  14 16 & +19 11 &  15.1 & 69.1  &  8 \\
5429 &                       & 3.58  &  K3III   & 45.6 &  14 32 &  +30 22 &   47.3 & 67.8  &  8 \\
5435 &                       & 3.00  &  A7III   & 26.1 &  14 32 &  +38 18 &   67.3 & 66.2  &  135 \\
5563 &                       & 2.08  &  K4III   & 38.8 &  14 51 &  +74 09 &  112.6 & 40.5  &  8 \\
5793 & $\alpha$ CrB          & 2.21  &  A0V     & 22.9 &  15 35 &  +26 43 &   41.9 & 53.8  &  132 \\
5849 &                       & 3.84  &  B9IV    & 44.5 &  15 43 &  +26 18 &   41.7 & 51.9  &  100 \\
5854 &                       & 2.64  &  K2IIIb  & 22.5 &  15 44 &  +06 26 &   14.2 & 44.1  &  8 \\
6092 &                       & 3.74  &  B5IV    & 96.4 &  16 20 &  +46 18 &   72.5 & 45.0  &  20 \\
6095 &                       & 3.74  &  A9III   & 59.9 &  16 22 &  +19 09 &   35.3 & 41.3  &  135 \\
6148 &                       & 2.79  &  G7IIIa  & 45.3 &  16 30 &  +21 29 &   39.0 & 40.2  &  10  \\
6149 &                       & 3.90  &  A0V     & 50.9 &  16 31 &  +01 59 &   17.1 & 31.8  &  142 \\
6212 &                       & 2.89  &  G0IV    & 10.8 &  16 41 &  +31 36 &   52.7 & 40.3  &  5 \\
6299 &                       & 3.20  &  K2III   & 26.3 &  16 58 &  +09 23 &   28.4 & 29.5  &  8 \\    \hline
6324 &                       & 3.91  &  A0V     & 49.9 &  17 00 &  +30 56 &   52.9 & 36.2  &  60 \\
6410 &                       & 3.13  &  A3IV    & 24.1 &  17 15 &  +24 50 &   46.8 & 31.4  &  305 \\
6556 & $\alpha$ Oph          & 2.10  &  A5III   & 14.3 &  17 35 &  +12 34 &   35.9 & 22.6  &  240 \\
6603 &                       & 2.77  &  K2III   & 25.1 &  17 43 &  +04 34 &   29.2 & 17.2  &  8 \\
6623 &                       & 3.42  &  G5IV    &  8.4 &  17 46 &  +27 43 &   52.4 & 25.6  &  8 \\
6629 &                       & 3.75  &  A0V     & 29.1 &  17 48 &  +02 42 &   28.0 & 15.4  &  212 \\
6688 &                       & 3.74  &  K2III   & 34.2 &  17 54 &  +56 52 &   85.2 & 30.2  &  8 \\
6703 &                       & 3.71  &  G8III   & 41.5 &  17 58 &  +29 15 &   54.9 & 23.8  &  10 \\
6705 &                       & 2.23  &  K5III   & 45.2 &  17 57 &  +51 29 &   79.1 & 29.2  &  8 \\
6872 &                       & 4.32  &  K2III   & 70.4 &  18 20 &  +36 04 &   63.5 & 21.5  &  8 \\
7001 & Vega, $\alpha$ Lyr    & 0.03  &  A0V     &  7.8 &  18 37 &  +38 47 &   67.4 & 19.2  &  5 \\
7235 &                       & 3.00  &  A0V     & 25.5 &  19 05 &  +13 52 &   46.9 &  3.2  &  360 \\
7405 & $\alpha$ Vul          & 4.45  &  M0III   & 90.9 &  19 29 &  +24 40 &   59.0 &  3.4  &   \\
7528 &                       & 2.90  &  B9.5IV  & 52.4 &  19 45 &  +45 08 &   78.7 & 10.2  &  128 \\
7557 & Altair, $\alpha$ Aql  & 0.77  &  A7V     &  5.1 &  19 51 &  +08 52 &   47.7 & $-$8.9 &  245 \\
7582 &                       & 3.83  &  G8III   & 44.6 &  19 48 &  +70 16 &  102.4 & 20.8  &   10 \\
7635 &                       & 3.53  &  M0III   & 84.0 &  19 59 &  +19 30 &   58.0 & $-$5.2 &  8 \\
\hline
\end{tabular}

\end{flushleft}
\label{tab_stars}
\end{table*}

\begin{table*}
\caption{Previous Polarization Measurements}
\begin{flushleft}
\begin{tabular}{lcrrrr}
\hline
BS   & Polarization (Heiles) & \multicolumn{2}{c}{Polarization (Tinbergen)$^a$} & \multicolumn{2}{c}{Polarization (Piirola)$^a$}\\
       &  \multicolumn{1}{c}{P(\%)}  &    Q/I  &  U/I  &  Q/I  &  U/I  \\   \hline                
3982 &  0.060$\pm$0.120   &  0$\pm$7   &  -4$\pm$7  &   7$\pm$11  &  3$\pm$  11\\
4031 &  0.050$\pm$0.120   &  &  &  &  \\
4295 &  0.000$\pm$0.120 &    4$\pm$7  &   6$\pm$7  &   13$\pm$6  &   3$\pm$6  \\
4301 &  0.040$\pm$0.120 &    3$\pm$7  &   22$\pm$7 &   $-$3$\pm$13  &  $-$2$\pm$13 \\
4335 &  0.030$\pm$0.120 &    22$\pm$7 &   $-$14$\pm$7   &   &\\
4357 &  0.000$\pm$0.120 &   &       &                       4$\pm$7   &    $-$6$\pm$7  \\
4359 &  0.010$\pm$0.120 &    $-$1$\pm$7 & $-$11$\pm$7   &    &   \\
4518 &  0.060$\pm$0.120 &    $-$33$\pm$7 & 1$\pm$7  &    &   \\
4534 &  0.030$\pm$0.120 &    12$\pm$7  &  3$\pm$7    &   0$\pm$14  &   $-$12$\pm$14 \\
4540 &  0.042$\pm$0.026 &    4$\pm$7   &  3$\pm$7   &   &  \\
4905 &  0.010$\pm$0.120 &    &    &    &  \\
4910 &  0.020$\pm$0.120 &    &    &    &  \\
4915 &  0.020$\pm$0.120 &    &    &    &   \\
4932 &  0.010$\pm$0.120 &    6$\pm$7   &  $-$6$\pm$7  &   &  \\
5191 &  0.060$\pm$0.000 &    &   &   &  \\
5235 &  0.007$\pm$0.012 &    $-$6$\pm$7  &  $-$12$\pm$7   &   $-$6$\pm$9  &  3$\pm$9  \\
5340 &  0.030$\pm$0.120 &    $-$1$\pm$7  &  $-$6$\pm$7  &   $-$10$\pm$8   &  $-$11$\pm$8  \\
5429 &  0.030$\pm$0.120 &    &    &   &  \\
5435 &  0.000$\pm$0.120 &    &    &   &  \\
5563 &  0.100$\pm$0.120 &    6$\pm$7   &  $-$1$\pm$7 &  &   \\
5793 &  0.060$\pm$0.120 &    &    &    14$\pm$6  &   $-$1$\pm$6  \\
5849 &  0.030$\pm$0.120 &    &    &   &  \\
5854 &  0.030$\pm$0.120 &    1$\pm$7   &  0$\pm$7  &  &  \\
6092 &  0.010$\pm$0.000 &    &    &   &  \\
6095 &  0.057$\pm$0.012 &    &    &   &  \\
6149 &  0.010$\pm$0.120 &    &   &   &  \\
6212 &  0.000$\pm$0.200 &    &   &   &   \\
6410 &  0.020$\pm$0.120 &    &   &   &   \\
6556 &  0.010$\pm$0.100 &    $-$4$\pm$7  &  $-$6$\pm$6  &  & \\
6603 &  0.150$\pm$0.120 &    &   &   &   \\
6629 &  0.008$\pm$0.001 &    12$\pm$7   &  15$\pm$7   &   &  \\
6688 &  &                 $-$36$\pm$7 & $-$41$\pm$7   &   &  \\
7001 &  0.020$\pm$0.120 &    11$\pm$7   &   7$\pm$7   &  4$\pm$4  &  6$\pm$4   \\
7528 &  0.030$\pm$0.120 &    &   &  &  \\
7557 & 0.016$\pm$0.002 &    15$\pm$7     &   0$\pm$7  &   2$\pm$6  &  $-$7$\pm$6   \\
\hline
\end{tabular}

a - Polarizations in units of 10$^{-5}$
\end{flushleft}
\label{tab_ppol}
\end{table*}

\section{Observations}

\subsection{The Sample Stars}

Stars selected for observation were in the RA range 10 to 20 hours, north of the equator, had V
magnitude brighter than 4.0, and were at a distance of less than 100 pc. 71 stars met these
criteria and 46 of them have been observed. In addition three stars were observed from a
supplementary list with a V magnitude limit of 5.0. The stars observed are listed in table
\ref{tab_stars}. This table gives the V magnitude and spectral type as listed in the SIMBAD
database, the distance derived from the Hipparcos catalogue parallax \citep{perryman97}, the approximate equatorial and
galactic coordinates (also from the Hipparcos catalogue) and the $v\sin{i}$ value, normally from 
\cite{bernacca70}, but in a few cases from other sources listed in the SIMBAD database. 
 Previous polarization measurements for the stars are
listed in table \ref{tab_ppol} have been taken from
the agglomerated polarization catalogue of \citet{heiles00}. The measurements of these stars from
\citet{heiles00} mostly originate from the work of \citet{behr59} (all those with errors of 0.12\%)
with a few measurements from other sources \citep{schmidt68,klare77,markkanen79}. Only the degree of
polarization is listed. The position angle can be found in the original catalogue, but for almost
all the measurements the polarization is not significant and the position angle is therefore
meaningless.

The \citet{heiles00} catalogue does not include some of the most accurate previous polarization
measurements of nearby stars made by \citet{piirola77} and \citet{tinbergen82}. These studies include a
number of the stars in our sample and are therefore listed separately in table \ref{tab_ppol}. The
measurements are in units of 10$^{-5}$ so need to be multiplied by 10 to be compared with the
observations reported here. The measurements of Tinbergen were made in three colour bands. We
have used, where available, the averaged band I and II measurements, and in other cases the band
I measurements. Almost all the previous polarization measurements do not show any
significant polarization     

\subsection{Observation Methods}

The observations were obtained with the PlanetPol polarimeter \citep{hough06} mounted on the 4.2m
aperture William Herschel Telescope (WHT) which is  located at the Observatorio de Roque de Los
Muchachos (ORM) at La  Palma in the Canary Islands. PlanetPol  achieves its high sensitivity
through the use of rapid modulation (40kHz) using   Photo-Elastic Modulators (PEMs). A three-wedge
Wollaston prism is used as the analyser and  beam splitter and the light is detected by two
avalanche photodiode (APD) detectors. A second  identical channel with its own PEM and APDs
monitors the sky background. 

The polarization measurements are made in a very broad red band covering wavelengths from 
590nm to 1000nm. The broad band is necessary to maximise the photon flux in order to
achieve  the high polarization sensitivity ($\sim 10^{12}$ photons are needed to reduce
photon shot noise  sufficiently to measure polarization levels of $\sim10^{-6}$). With such a
broad band the precise effective wavelength depends on the colour of the star, and ranges from
735 nm for a B0 V star to 804.4 nm for an M5 V star. Full details are given in table 1 of
\citet{hough06}.

A number
of corrections are needed to  achieve the $10^{-6}$ sensitivity (as described in more
detail by \citet{hough06}). All observations use a ``second-stage chopping'' procedure in
which  a periodic rotation of the detectors and Wollaston prism through 90 degrees
relative to the PEM  is carried out to reverse the sign of the modulation. This was
normally done after every 180 seconds of integration. The telescope introduces a small
polarization ($\sim10-20 \times 10^{-6}$ for the  WHT). This telescope polarization is
determined by carrying out observations of stars repeated at  several different hour
angles. For an altazimuth-mounted telescope like the WHT, the telescope tube  rotates
about its optical axis relative to the sky (and the polarimeter which is mounted on a 
rotator that tracks the sky), as the telescope tracks, and this allows the telescope
polarization to  be separated from the star polarization. A small instrumental
polarization $\sim2 \times 10^{-6}$, (believed to  be due to misalignments in the
instrument) is determined by repeating measurements with the  entire instrument rotated
through 90 degrees. Separate measurements with the instrument rotated  through 45 degrees
are used to determine the Q and U Stokes parameters. These can then be  combined to give
the degree of polarization and the position angle. Measurements of a number of  stars with
known large polarizations ($\sim$1 -- 5 \%) are used to determine the instrument's 
modulation efficiency, which is found to be 98.6\% after the known effects of the PEM are allowed
for. The same observations are used to determine the position angle zero point which is measured with an accuracy of 1
degree, based on the consistency of the calibration provided by different reference stars, which themselves have
uncertainties of this order. 

The bulk of the measurements described here were obtained over the period 25 April 2005 to 9 May 
2005. Most stars were observed once, using a total integration time of usually 12 minutes for each Stokes
parameter. Four stars (BS 4932, BS 5854, BS 5445, BS4534) were observed repeatedly as calibrators
for the telescope polarization and the polarizations have been averaged over 4 to 6 observations.
The individual measurements for these stars are given by \citet{bailey08}. Two stars (BS 4295 and
BS 4540) were similarly observed as calibrators during February 2006 and the individual
observations can be found in \citet{lucas09}. BS 5793 was observed three times and the three
observations have been averaged. BS 3982 was also observed three times and the average of the two
best observations has been used. BS 4031 and BS 7001 were both observed twice, but one
observation was in significantly better conditions, so only the better of the two observations was
used.

\subsection{Saharan Dust Correction}

As explained by \citet{bailey08} and \citet{ulanowski07} the nights of May 3 to 7 2005 were
affected by an airborne Saharan dust event which introduced a small spurious polarization in the
horizontal direction. The excess polariation was negligible at the zenith but increased
with zenith distance. Observations on these nights have been handled as follows:

\begin{enumerate}
\item Observations on May 3, the most dust affected night, were not used.
\item On May 4 to 7 observations were only used if no other observations for the
star were available, and if they were made near the zenith where the
effects of the Saharan dust are minimal. All observations were at zenith distance less than
15 degrees on May 4 and less than 20 degrees on May 5-7.
\item A correction was applied to these measurements for the dust induced polarization.
\item The estimated errors for these observations were increased by adding half the dust
correction in quadrature to allow for uncertainties in the precise zenith distance
dependence of the dust induced polarization.
\end{enumerate}
The correction for the dust induced polarization has the form:

\begin{equation}
P_{dust} = 1.19 \times 10^{-4} (1 - \cos{ZD})
\end{equation}

for May 4, with the size of the correction on the subsequent nights reduced in proportion
to the dust optical depth on these nights. This was found to provide a reasonable fit to
the dust affected observations reported by \citet{bailey08}. The size of the correction was
at most 4 $\times$ 10$^{-6}$, and only four observations required a correction of more
than 2 $\times$ 10$^{-6}$.

\begin{table*}
\caption{Planetpol Linear Polarization Measurements - The horizontal line marks the 17 hour division used in figure 4}
\begin{flushleft}
\begin{tabular}{llrrrr}
\hline
Star & Date(s)$^a$ & \multicolumn{1}{c}{Q/I$^{b}$} & 
       \multicolumn{1}{c}{U/I$^{b}$} & 
       \multicolumn{1}{c}{P$^{b,c}$} &
       \multicolumn{1}{c}{$\theta$$^{b}$} \\
\hline
BS 3982 & Apr 26,Feb 20 & $-$34.0$\pm$0.8 &    13.8$\pm$0.8 &  36.7$\pm$0.8 &  78.9$\pm$0.8 \\
BS 4031 & Apr 29  &  $-$11.6$\pm$2.4 &  $-$8.0$\pm$2.4 &  14.1$\pm$2.5 & 107.3$\pm$4.9 \\
BS 4069 & Apr 27  &   $-$7.5$\pm$1.1 & $-$17.4$\pm$1.1 &  18.9$\pm$1.1 & 123.4$\pm$1.6 \\
BS 4295 & Feb 16-21 &    5.0$\pm$0.9 &  $-$8.2$\pm$1.0 &   9.6$\pm$1.0 & 150.7$\pm$4.2 \\
BS 4301 & Apr 25  &      2.1$\pm$0.9 &  $-$9.1$\pm$0.9 &   9.3$\pm$0.9 & 141.6$\pm$2.7 \\
BS 4335 & Apr 30  &   $-$3.0$\pm$2.1 &  $-$2.4$\pm$2.1 &   3.8$\pm$2.1 & 109.5$\pm$15.3 \\
BS 4357 & Apr 30  &      2.7$\pm$2.2 &  $-$2.5$\pm$2.5 &   3.7$\pm$2.4 & 158.5$\pm$18.3 \\
BS 4359 & May 5   &      4.4$\pm$2.7 &     5.3$\pm$2.7 &   6.9$\pm$2.7 &  25.2$\pm$11.3 \\ 
BS 4518 & May 6   &      4.7$\pm$2.6 &  $-$8.9$\pm$2.6 &  10.1$\pm$2.6 & 148.9$\pm$7.3 \\
BS 4527 & May 7   &  $-$10.9$\pm$3.7 &  $-$4.2$\pm$3.6 &  11.8$\pm$3.7 & 100.6$\pm$8.8 \\
BS 4534 & Apr 27-30 &    0.8$\pm$1.1 &     2.2$\pm$1.1 &   2.3$\pm$1.1 &  35.3$\pm$13.9 \\
BS 4540 & Feb 18-21 &    3.3$\pm$1.4 &  $-$0.1$\pm$1.4 &   3.3$\pm$1.4 & 178.9$\pm$10.0 \\
BS 4905 & Apr 26    &   16.5$\pm$1.2 &     0.5$\pm$1.3 &  16.5$\pm$1.2 &   0.8$\pm$2.3 \\
BS 4910 & Apr 28    &    1.5$\pm$1.6 &  $-$2.5$\pm$1.5 &   2.9$\pm$1.5 & 150.8$\pm$15.4 \\
BS 4915 & May 4     & $-$7.8$\pm$2.6 &  $-$4.2$\pm$2.3 &   8.8$\pm$2.6 & 104.1$\pm$7.6 \\
BS 4932 & Apr 27-29, May 8 &  5.6$\pm$1.0 &     0.6$\pm$1.0 &   5.6$\pm$1.0 &   3.0$\pm$3.9 \\
BS 5054 & May 8     &    7.4$\pm$1.6 &  $-$1.8$\pm$1.5 &   7.6$\pm$1.6 & 173.3$\pm$5.8 \\
BS 5191 & Apr 25    &    9.3$\pm$1.7 &  $-$2.4$\pm$2.1 &   9.7$\pm$1.7 & 172.7$\pm$6.1 \\
BS 5235 & Apr 27    &    3.2$\pm$1.8 &  $-$1.5$\pm$1.5 &   3.5$\pm$1.8 & 167.3$\pm$12.9 \\
BS 5340 & May 9     &    3.0$\pm$1.4 &     5.5$\pm$1.6 &   6.3$\pm$1.6 &  30.6$\pm$6.7 \\
BS 5429 & May 4     & $-$3.7$\pm$2.2 &  $-$9.6$\pm$2.5 &  10.3$\pm$2.4 & 124.5$\pm$6.2 \\
BS 5435 & Apr 27-30 & $-$2.8$\pm$1.6 &  $-$2.2$\pm$1.6 &   3.6$\pm$1.6 & 108.6$\pm$9.8 \\ 
BS 5563 & May 9     &    8.2$\pm$2.2 &     1.9$\pm$1.8 &   8.4$\pm$2.2 &   6.5$\pm$6.3 \\
BS 5793 & Apr 25-26 & $-$3.9$\pm$1.2 &     0.0$\pm$1.0 &   3.9$\pm$1.2 &  90.0$\pm$7.2 \\
BS 5849 & May 7     & $-$0.4$\pm$3.3 &     0.0$\pm$3.7 &   0.4$\pm$3.3 &  90.7$\pm$235 \\ 
BS 5854 & Apr 27-30, May 8 & $-$2.3$\pm$0.9 &     3.9$\pm$1.0 &   4.5$\pm$0.9 &  59.7$\pm$6.1 \\
BS 6092 & Apr 28  & $-$234.6$\pm$3.3 & $-$19.7$\pm$3.4 & 235.4$\pm$3.3 &  92.3$\pm$0.4 \\
BS 6095 & May 5   &   $-$1.3$\pm$2.5 &    13.2$\pm$2.9 &  13.3$\pm$2.9 &  47.9$\pm$5.3 \\
BS 6148 & Apr 26    &   12.3$\pm$1.4 &    14.1$\pm$1.3 &  18.7$\pm$1.3 &  24.5$\pm$2.1 \\
BS 6149 & Apr 29    &   10.6$\pm$3.3 &     3.6$\pm$3.0 &  11.2$\pm$3.3 &   9.4$\pm$7.6 \\
BS 6212 & May 4     & $-$2.2$\pm$2.6 &     9.3$\pm$2.6 &   9.6$\pm$2.6 &  51.6$\pm$7.8 \\
BS 6299 & Apr 29    & $-$7.4$\pm$1.7 &     9.4$\pm$1.4 &  11.9$\pm$1.5 &  64.2$\pm$3.9 \\  \hline
BS 6324 & May 4     & $-$8.3$\pm$4.1 &     9.2$\pm$3.7 &  12.4$\pm$3.9 &  65.9$\pm$9.1 \\ 
BS 6410 & May 5     & $-$5.3$\pm$2.4 &     5.8$\pm$2.4 &   7.9$\pm$2.4 &  66.1$\pm$8.8 \\
BS 6556 & Apr 27    &   11.1$\pm$2.0 &    20.6$\pm$2.0 &  23.4$\pm$2.0 &  30.8$\pm$2.4 \\
BS 6603 & Apr 27    &   18.1$\pm$2.1 &    26.6$\pm$2.2 &  32.2$\pm$2.1 &  27.9$\pm$1.9 \\
BS 6623 & May 6     &    6.9$\pm$2.2 &     6.2$\pm$1.9 &   9.3$\pm$2.1 &  20.9$\pm$6.2 \\
BS 6629 & May 8     &   22.2$\pm$3.0 &    34.3$\pm$3.2 &  40.8$\pm$3.1 &  28.5$\pm$2.1 \\
BS 6688 & Apr 28    & $-$0.8$\pm$3.3 &     3.7$\pm$2.9 &   3.8$\pm$2.9 &  50.8$\pm$25.3 \\ 
BS 6703 & May 4     &   22.8$\pm$2.4 &     8.4$\pm$2.4 &  24.3$\pm$2.4 &  10.1$\pm$2.8 \\
BS 6705 & Apr 28    &   25.1$\pm$1.2 & $-$68.8$\pm$1.2 &  73.3$\pm$1.2 & 145.0$\pm$0.5 \\
BS 6872 & May 7     &  104.1$\pm$2.7 &    22.5$\pm$2.7 & 106.5$\pm$2.7 &   6.1$\pm$0.7 \\
BS 7001 & May 6     &    6.2$\pm$1.0 &    16.1$\pm$1.0 &  17.2$\pm$1.0 &  34.5$\pm$1.4 \\
BS 7235 & May 4   &  $-$13.0$\pm$3.0 &    18.7$\pm$3.0 &  22.8$\pm$3.0 &  62.4$\pm$3.8 \\
BS 7405 & May 7     &  175.3$\pm$2.3 &  1309.7$\pm$3.1 & 1321.4$\pm$3.1 & 41.2$\pm$0.1 \\
BS 7528 & May 8     &   75.6$\pm$2.0 &    77.0$\pm$2.1 & 108.0$\pm$2.0 &  22.8$\pm$0.5 \\
BS 7557 & Apr 26    & $-$7.3$\pm$1.3 &  $-$1.2$\pm$1.2 &   7.4$\pm$1.3 &  94.6$\pm$4.7 \\
BS 7582 & May 8     & $-$0.9$\pm$2.7 &     1.2$\pm$2.5 &   1.5$\pm$2.6 &  62.8$\pm$50.3 \\
BS 7635 & Apr 26   & $-$75.1$\pm$1.5 &   184.9$\pm$1.4 & 199.5$\pm$1.4 &  56.1$\pm$0.2 \\
\hline
\end{tabular}

a - April and May dates are 2005, February dates are 2006 \\
b - All polarizations are in units of 10$^{-6}$, position angles
are in degrees. \\
c - P is calculated from $P = \sqrt{(Q/I)^2 + (U/I)^2}$ 
\end{flushleft}
\label{tab_results}
\end{table*}

\subsection{Results}

The resulting polarization measurements for the 49 stars are listed in table
\ref{tab_results}. This table lists the normalized Stokes parameters Q/I and U/I, which are
obtained in separate Planetpol observations, and the degree of polarization and position
angle obtained by combining the Q/I and U/I measurements. The polarizations and Stokes
parameters are in units of 10$^{-6}$ in fractional polarization.

The errors quoted are derived from the internal statistics of the individual data points included in
each measurement as described by \cite{hough06} and include the uncertainties in the determination of
the telescope polarization. As discussed in \cite{hough06} and \cite{lucas09}, analysis of stars with
repeated observations suggest that this procedure may somewhat underestimate the true errors (by
factors up to 1.8) for the brighter stars in the sample where internal errors of $\sim$1 $\times$
10$^{-6}$ or better are achieved. The quoted errors are probably a better measurement of the true
uncertainty for the fainter stars in the sample.

\begin{figure}
\includegraphics[width=84mm]{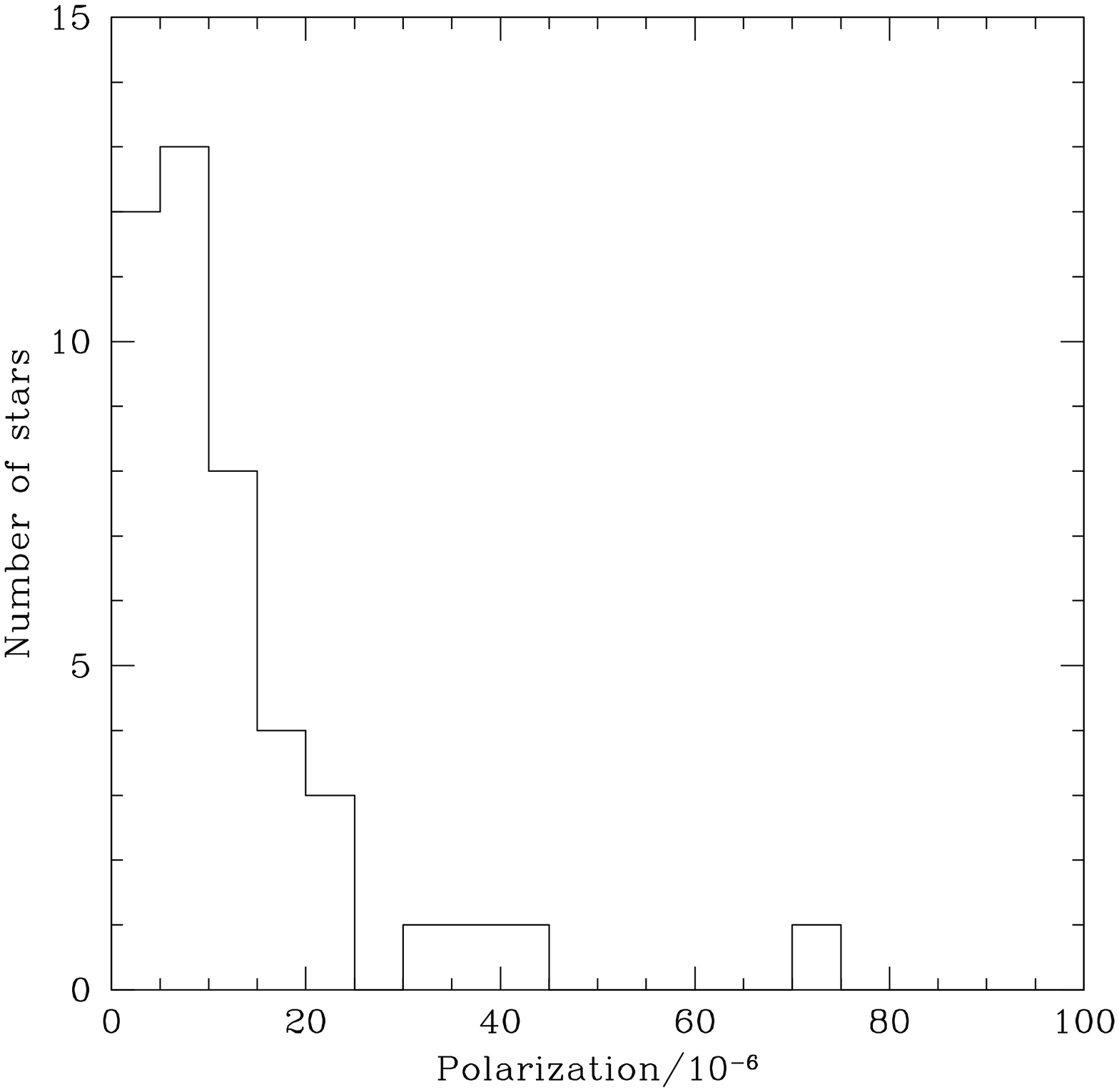}
\caption{Histogram of polarizations of stars in the sample. There are 5 stars in the
sample with P $>$ 10$^{-4}$ and therefore not shown here.}
\label{fig_hist}
\end{figure}

\begin{figure}
\includegraphics[width=84mm]{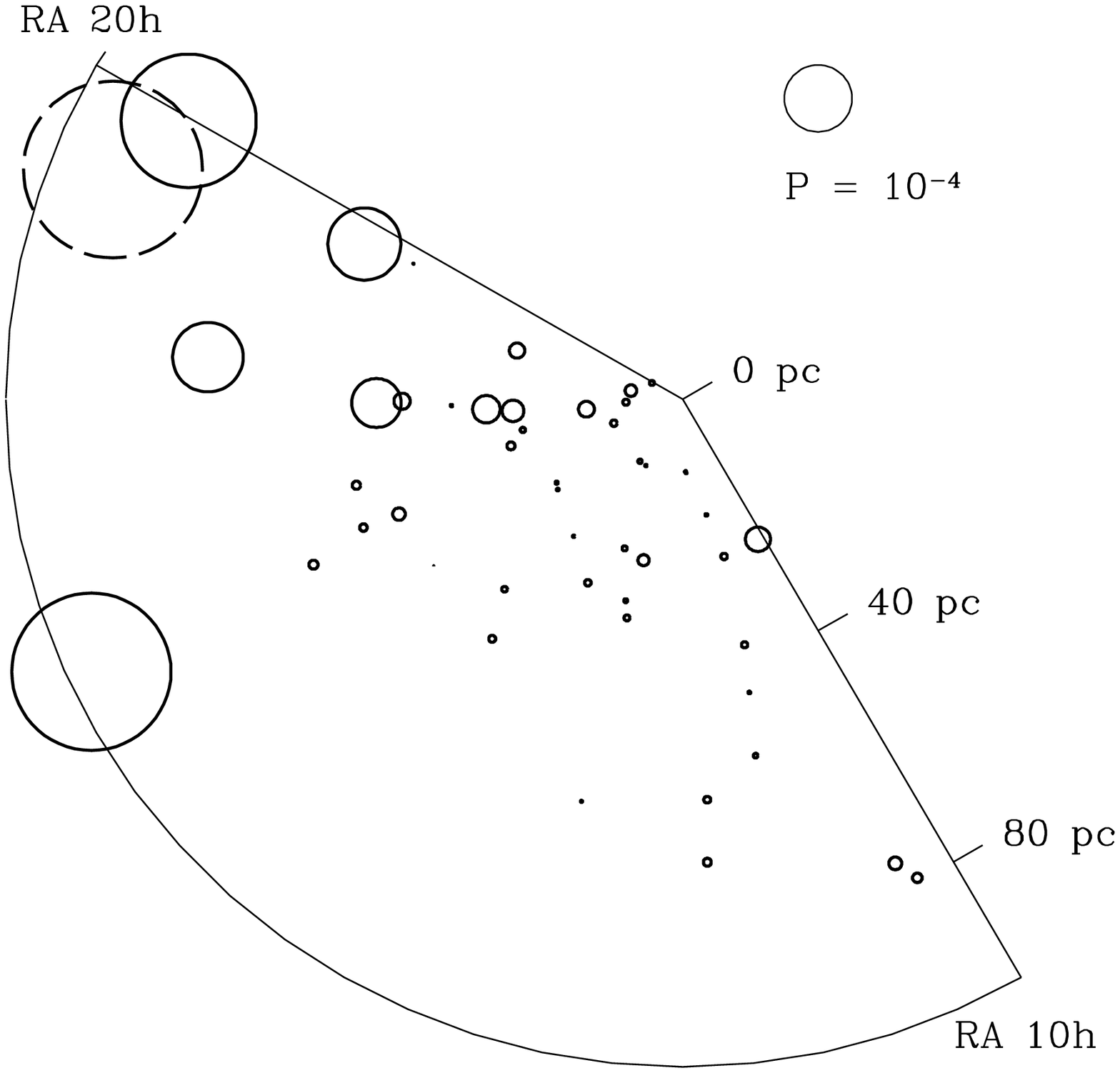}
\caption{Spatial distribution of polarization in the sample. The size of the circle is
proportional to the degree of polarization, except for the dashed circle which represents
BS7405 and is shown 5 times smaller than its correct size.}
\label{fig_pos}
\end{figure}

\section{Discussion}

\subsection{Polarization Statistics and Previous Observations}

While previous polarization studies of nearby stars have shown very few stars with
significant polarization, the much increased sensitivity of our study reveals polarization
to be much more common, with 38 stars (77.6\% of the sample) having polarizations that exceed
3-sigma. Figure \ref{fig_hist} shows a histogram of the measured polarizations and this
shows that 24 stars (48.9\% of the sample) have polarization above 10$^{-5}$. Most of the
polarizations are, however, relatively small. Only 9 stars in the sample have polarizations
above 2.5 $\times$ 10$^{-5}$, and only two exceed 2 $\times$ 10$^{-4}$. 

Hence it is not surprising that previous surveys of the polarization of nearby stars, with
precisions of at best $\sim$7 $\times$ 10$^{-5}$ (\cite{tinbergen82} and \cite{piirola77}),
detected few polarized stars.
For these surveys a 3-sigma detection would require a polarization of at least 2.1 $\times$
10$^{-4}$, and only BS 7405 in our sample has a large enough polarization to be clearly
detected. However, this star, and our two next highest polarizations BS 6092 and BS 7635 
were not observed in either of these previous studies.

\citet{tinbergen82} did report marginally significant polarizations in three of our sample
stars, BS 4335, BS 4518 and BS 6688. However, all these stars show very low polarization in
our observations.  

In table \ref{tab_spc} the polarization properties as a function of spectral type are listed.
Across the spectral classes A, F, G and K the median polarizations, and the percentage of stars
with polarization greater than 15 $\times$ 10$^{-6}$ are very similar. \cite{tinbergen82} suspected
the presence of variable intrinsic polarization at the 10$^{-4}$ level in stars with spectral 
type F0 and later. This is not supported by our more sensitive observations. Higher polarizations
are indicated for spectral classes B and M, but we have very few stars of these types, and they are
at larger average distances, so this result is almost certainly a consequence of the polarization
distance relation we find in the next section, and not a consequence of any intrinsic polarization of
these stars.

\begin{table}
\caption{Polarization for spectral types}
\begin{tabular}{lllrl}
\hline
Spectral & N  &  Mean  &  Median & Percent $>$  \\
Type  &      &  Dist  &  Poln   & 15 $\times$ 10$^{-6}$ \\  \hline
B      &   5  &  49.6  &  36.8 $\times$ 10$^{-6}$   & 60    \\
A      &   18 &  28.1  &  8.4  $\times$ 10$^{-6}$  & 28    \\
F/G    &   9  &  23.5  &  9.4  $\times$ 10$^{-6}$  & 22    \\
K      &   12 &  38.7  &  9.8  $\times$ 10$^{-6}$  & 25    \\
M      &   4  &  78.6  &  109.2 $\times$ 10$^{-6}$ & 75    \\
\hline
\end{tabular}
\label{tab_spc}
\end{table}

\subsection{Polarization Spatial Distribution}

\begin{figure*}
\includegraphics[width=160mm,viewport=0 70 600 450,clip]{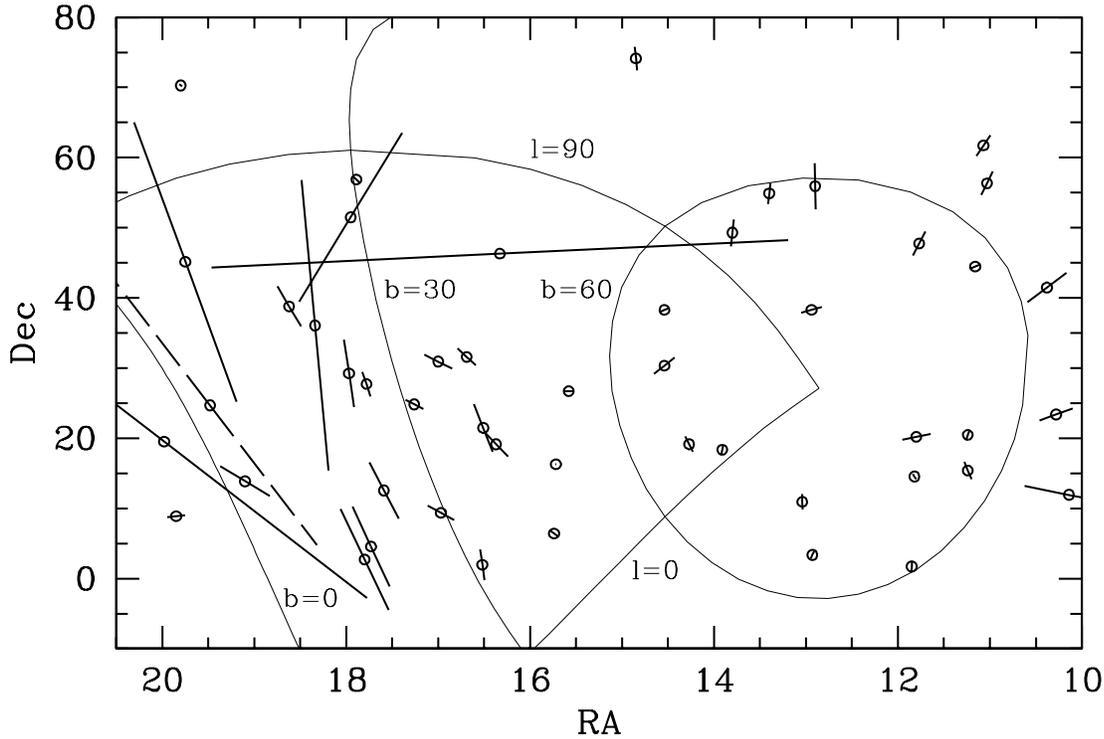}
\caption{Polariation vectors plotted against position. Lines of galactic
latitude 0, 30 and 60, and galactic longitude 0 and 90 are shown. The dashed vector
represents BS7405 and is shown 10 times smaller than its correct size.}
\label{fig_radec}
\end{figure*}

The spatial distribution of the observed polarizations is shown in figures \ref{fig_pos}
and \ref{fig_radec}. The distribution is quite striking. Figure \ref{fig_pos} shows that
there is very little polarization in the RA range 10 to 16 hours, and that most of the high
polarizations are found in the range 16 to 20 hours. Within this RA range there is a strong
correlation with distance, with the highest polarizations being found at the greatest
distances.

In figure \ref{fig_radec} it can be seen that the high polarizations correspond to lower
galactic latitudes. Stars around the galactic pole have generally low polarizations and
substantial polarizations begin to appear for galactic latitudes below about 40 degrees. It
is also apparent from the polarization vectors that the position angles are not random, but
show a tendency for the polarization direction to be along lines of galactic latitude. This implies a
galactic magnetic field in the solar vicinity that lies in the galactic plane.
These correlations with position strongly suggest that the bulk of the polarization being
observed is interstellar in origin, and not intrinsic to the stars. 

\begin{figure}
\includegraphics[width=84mm]{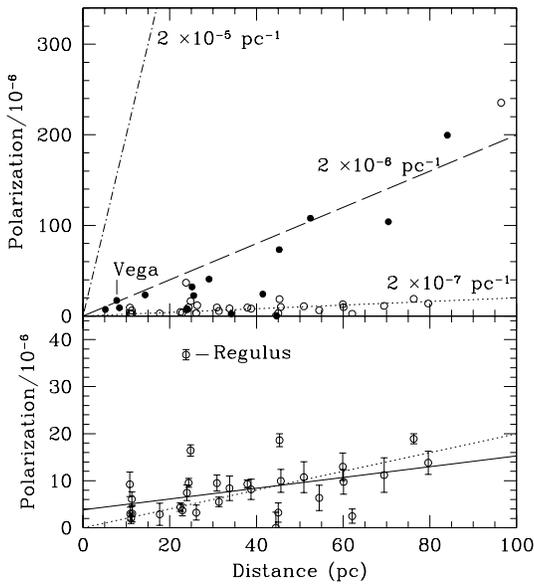}
\caption{Polarization plotted against distance. Solid symbols are stars with RA $>$ 17h
and open symbols are stars with RA $<$ 17h. The dot-dash line shows the polarization
versus distance relation for distant stars and the dash and dotted lines are relationships
with 10 and a 100 times less polarization per parsec. The lower panel is an expanded view 
of the data for stars with RA $<$17h. The solid line is the least squares fit to the data
excluding Regulus and BS 6092}
\label{fig_dist}
\end{figure}

Figure \ref{fig_dist} shows the variation of polarization with distance. Studies of the interstellar
polarization of more distant stars have shown that fractional polarization increases with distance at about
2 $\times$ 10$^{-5}$ pc$^{-1}$ \citep{behr59}. The studies of \citet{tinbergen82} and \citet{leroy93b} have shown that 
the polarization of nearby stars is less than would be expected from this relationship. Figure
\ref{fig_dist} shows just how low polarizations near the Sun are compared with this relationship.
The open circles on this figure, which represent stars at RA less than 17h and corresponds to regions
around the north galactic pole, actually fit a relationship
of about 2 $\times$ 10$^{-7}$ pc$^{-1}$, about 100 times less than that for distant stars. This
indicates that the interstellar medium in this direction has about a factor of 100 less dust than the
typical value for the interstellar medium in the galactic plane. 

The lower panel of figure \ref{fig_dist} shows an expanded view of the polarization data for
RA $<$17h. The polarization data used in figure 4 have been debiased by 
plotting $\sqrt{P^2 - \sigma_P^2}$ where $\sigma_P$ is the error in 
polarization given in table \ref{tab_results}.
This is a standard method of correcting for the fact that polarization values are always positive and can
therefore be significantly biased toward larger values where P is comparable with $\sigma_P$.
Even at these low polarization levels there is an indication that polarization increases with
distance. The solid line in the lower panel of figure 4 is the least squares fit to the data (excluding
Regulus and BS 6092). The correlation coefficient is 0.469 and the probability of this occurring
by chance is less than 2\%. This demonstrates that even in this low polarization region we are seeing
a contribution from interstellar polarization. It does not rule out small amounts of polarization
from other sources, but neither are other mechanisms required. An imperfect correlation with distance
is expected due to the patchy nature of the density of the interstellar medium.

In the region nearer the galactic plane shown by the solid symbols in figure \ref{fig_dist} the
polarizations are higher and some measurements fit an increase with distance of about 
2 $\times$ 10$^{-6}$ pc$^{-1}$ about a factor 10 less than the value for distant stars. However, many
points fall well below this line suggesting that the dust in this region is very clumpy.

Polarization is related to the amount of interstellar dust along the line of sight. However, the
degree of polarization will also depend on the efficiency of grain alignment, and on the angle of the
magnetic field to the line of sight. The relationship between polarization and extinction 
 $E(B-V)$ for more distant stars, shows that the maximum polarization is given by $P(\%) \sim
9 E(B-V)$ \citep{schmidt68}, corresponding to maximum grain alignment, but actual values show a great
deal of scatter and can fall well below this maximum value. \cite{fosalba02} give a mean relationship
of $P(\%) = 3.5 E(B-V)^{0.8}$. 

If we apply this relationship to the two lower lines shown on figure \ref{fig_dist}, then the lines
correspond to a $E(B-V)$ = 0.00157 at 100 pc for the line that roughly fits the galactic plane
polarizations, and $E(B-V)$ = 0.000037 at 100 pc for the lower line that fits the galactic pole polarizations.
These $E(B-V)$ values are well below those that can actually be measured by photometry.

The relationship between polarization and angle to the line of sight should lead to a polarization that depends on
galactic longitude. This is difficult to test with our data, since the stars near the galactic plane are restricted to a
small range of galactic longitude. A more extensive survey, might reveal such effects.

\subsection{Polarization and the Local Cavity}

The distribution of polarization we observe is consistent with other data on the local
interstellar medium that shows the presence of a local cavity or local bubble in the solar
vicinity \citep{cox87,lallement07}. For example, studies of the 3D distribution of interstellar gas
using Na I D line absorption \citep{lallement03}, show little interstellar gas near the Sun and
towards the north galactic pole, but show that significant gas is detected quite close to the Sun
($\sim$50 pc) in the galactic plane between galactic longitudes 0 and 90, the direction in which
we see the largest polarizations.  This suggests that the distribution of dust shown by the
polarization is similar to the gas distribution revealed by the Na I absorption measurements. A
similar distribution of interstellar material is shown in the local hot bubble (LHB) revealed in
soft X-ray background observations \citep{snowden98} which appears to lie within the cavity seen
in Na I absorption.

Our polarization observations confirm the presence of a Local Cavity with little interstellar
material within 100pc of the Sun, but nevertheless suggest that there is some interstellar dust 
within this region that can be detected by high-sensitivity polarimetry. It is less clear whether
the polarization observations define a sharp edge to the local cavity. The large polarizations 
observed for BS 7405 and BS 6092, two of the most distant stars in our sample, may be indicative of
such an edge, but there are too few stars at this distance to determine if such polarizations are
typical of this distance ($\sim$90 pc).

We are also unable to directly look at correlations between polarization
and sodium absorption on an individual sight line basis, as there is only one star common
to our sample and the Na I observations of \citet{lallement03} and \citet{sfeir99}, and
this star has both small polarization and no significant D line absorption.

\begin{figure}
\includegraphics[width=84mm]{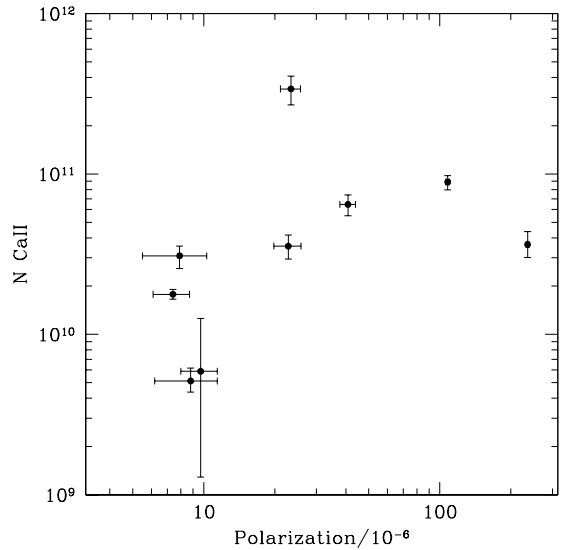}
\caption{Ca II column density plotted against polarization for nine stars}
\label{fig_ca2}
\end{figure}

While we cannot directly compare with Na I D line absorption, there are a number of measurements
of Ca II H\&K absorption for nearby early-type stars included in our sample. Ca II absorption is
seen in sight lines towards nearby stars that do not show significant neutral gas as seen in Na I
absorption. The Ca II absorption is thought to result from warmer gas present in the local cavity.
In many cases there are several distinct velocity components seen in the CaII absorption indicating
the presence of discrete clouds.
In figure \ref{fig_ca2} we show Ca II column density for nine stars from our sample. The CaII data
are taken from \cite{redfield02} and are originally from \cite{bertin93}, \cite{crawford98},
\cite{lallement92}, \cite{vallerga93} and \cite{welty96}. Where multiple
velocity components are observed, the column density has been summed over all components.
Figure \ref{fig_ca2} shows a slight tendency for higher Ca II column densities at higher
polarizations, but there is much scatter and no strong correlation. This suggests that the dust
responsible for the polarization is not located in the clouds responsible for the Ca II absorption.

\subsection{Individual Stars}

BS 3982 (Regulus) stands out as having a high polarization compared with stars in its RA
range despite being at a relatively small distance of 23.8pc. Two observations of its
polarization in 2005 April and 2006 February are in excellent agreement.
Regulus is known to be a rapidly
rotating B star, and its rotational flattening has been measured directly by
interferometry. \citet{mcalister05} used the CHARA array to determine the rotational
flattening of Regulus and derive a position angle of the minor axis of 85.5 $\pm$ 2.8
degrees. This is in agreement with our measured polarization position angle of 78.9
$\pm$0.8 degrees.

A rotationally flattened early type star is expected to show polarization as a result of
electron scattering in its non-spherically symmetric atmosphere. \citet{sonneborn82}
calculates the polarization  expected in rapidly rotating B stars. The polarization is
parallel to the rotation axis at red wavelengths as observed for Regulus. The degree of
polarization decreases with spectral subclass from about 0.01\% (100 $\times$ 10$^{-6}$) at
B2 to about 0.005\% (50 $\times$ 10$^{-6}$) at B5, the latest spectral class modelled. Our
measured polarization for Regulus (B7) of 36.7 $\times$ 10$^{-6}$ is therefore in good
agreement with these calculations. However, \citet{sonneborn82} calculates the polarization
for a star rotating with 95\% of its critical velocity for break-up, whereas according to
\citet{mcalister05} Regulus is only rotating at 86\% of its critical velocity. It is therefore
important to carry out polarization calculations for rapidly rotating stars 
that include a wider range of parameters. However, the fact that we see unusual polarization 
for Regulus, and this star stands out in our sample as having both rapid rotation and an
early spectral type (needed to give an electron scattering atmosphere), strongly suggests that 
the polarization arises from rotational flattening. If this interpretation is correct it would be the first observation
of this effect.

BS 7001 (Vega) shows a polarization of 17.2 $\pm$ 1.0 $\times$ 10$^{-6}$ at a position angle of
34.5 $\pm$ 1.4 degrees. Vega is known to possess a debris disk detected by its infrared excess
\citep{aumann84}. Scattering of light from dust in the disk is therefore a potential source of
polarization, as has been observed in the case of the Beta Pictoris disk \citep{gledhill91,tamura06}.
However, the Vega disk is large. Spitzer observations show it extending out to 105 arc seconds at
160 $\mu$m \citep{su05} and it is circular indicating that the disk is seen face-on. The Spitzer
observations also indicate an inner radius for the disk of 11 $\pm$ 2 arc seconds. Since the PlanetPol
observations used an aperture of 5 arc seconds this would mean that the observations did not include the
disk. However, there is also evidence for circumstellar material within 1 arc second of Vega
\citep{absil06}.  The face-on nature of the system is supported by interferometric observations of Vega that show it
to be a rapidly rotating star seen almost pole-on \citep{peterson06,aufdenberg06}, with an inclination
measured to be 4.7 $\pm$ 0.3 \citep{aufdenberg06} or 4.55 $\pm$ 0.33 degrees \citep{peterson06}.

There have been no detections of the Vega disk in scattered light. \citet{mauron98} attempted to detect
scattered light through polarization at distances of 7 to 30 arc seconds for the star, and detected no
polarization with limits about 200 times lower than the disk of Beta Pic. A face-on disk with material
symmetrically distributed around the star would not be expected to show any polarization in observations
centered on the star. Thus our observed polarization could only result from the circumstellar disk if
there is an asymmetric distribution of circumstellar material close to the star.

An alternative explanation of the Vega polarization is that it is interstellar in origin. This may seem
unlikely for a star at a distance of only 7.8 pc. However, Vega is located in the region of the sky
where we see the largest interstellar polarizations. The polarization would correspond to 2.2 $\times$
10$^{-6}$ pc$^{-1}$, which is similar to that seen in more distant stars in this RA range, as seen in
figure \ref{fig_dist}. The position angle of 35 degrees is similar to that of other stars in the region.

\subsection{Interstellar Dust in the Heliosphere}

\cite{frisch05} has argued that the weak polarization of nearby stars observed by
\cite{tinbergen82} could be due to interstellar dust entrained in the magnetic wall of the
heliosphere. This was based on the presence of a spatial distribution of polarization that was
related to ecliptic coordinates, and the absence of a significant distance dependence of the
polarization. Our results do not support this interpretation for polarization of nearby stars.
Figures \ref{fig_pos} and \ref{fig_dist} show clearly that polarization is correlated with 
distance and that the dust responsible must therefore be 
widely distributed over the 100 pc scale covered by our observations.

\section{Conclusions}

Polarization measurements of a sample of 49 nearby bright stars have been measured to
accuracies about 20 to 100 times better than those of any previous measurements. In contrast 
to previous observations which have generally been unable to detect many polarized stars at these
distances, we find significant polarization in many of the stars. The polarization increases with
distance and shows much higher values at low galactic latitudes than towards the galactic pole.
The distribution of polarization strongly suggests that the high polarization stars and probably
most of the lower polarization stars, are showing interstellar polarization. The results indicate
that polarization measured at the parts per million level provides
a very sensitive probe of the interstellar medium in the solar vicinity.

The polarization observed near the Sun is much less than would be expected based on the polarization of
distant stars, thus confirming the presence of the local cavity or bubble seen in absorption
line measurements and in the soft X-ray background. Polarization shows litle correlation with CaII
absorption due to warm interstellar gas. The data is not consistent with the hypothesis of
\cite{frisch05} that polarization in nearby stars is due to interstellar dust entrained in the heliospere. 

Regulus shows a larger polarization than expected for its position. Regulus is known to be a
rapidly rotating star, and the polarization direction agrees with the minor axis of the rotational
flattening as measured by interferometry. The polarization is reasonably consistent with that
expected due to electron scattering in the atmosphere of the flattened star.

\section*{Acknowledgments}

We acknowledge the award of a PPARC grant to build the PlanetPol polarimeter. We thank the staff
of the Isaac Newton Group for their support when using PlanetPol on the WHT. We thank
Edwin Hirst and David Harrison for their support of the PlanetPol instrument during
the observing runs.

\label{lastpage}

\end{document}